\documentclass[%
reprint,
superscriptaddress,
showpacs,
 amsmath,amssymb,
 aps,
]{revtex4-2}

\usepackage{graphicx}
\usepackage{dcolumn}
\usepackage{bm}

\begin{document}

\preprint{APS/123-QED}

\title{%
  First direct observation of isomeric decay in neutron-rich odd-odd $^{186}$Ta
}%

\author{Y.X.~Watanabe}%
\email{yutaka.watanabe@kek.jp}%
\affiliation{Wako Nuclear Science Center, Institute of Particle and Nuclear Studies, High Energy Accelerator Research Organization (KEK), Wako, Saitama 351-0198, Japan}%
\author{P.M.~Walker}%
\affiliation{Department of Physics, University of Surrey, Guildford, GU2 7XH, United Kingdom}%
\author{Y.~Hirayama}%
\affiliation{Wako Nuclear Science Center, Institute of Particle and Nuclear Studies, High Energy Accelerator Research Organization (KEK), Wako, Saitama 351-0198, Japan}%
\author{M.~Mukai}%
\affiliation{University of Tsukuba, Tsukuba, Ibaraki 305-0006, Japan}%
\affiliation{Wako Nuclear Science Center, Institute of Particle and Nuclear Studies, High Energy Accelerator Research Organization (KEK), Wako, Saitama 351-0198, Japan}%
\affiliation{RIKEN Nishina Center for Accelerator-Based Science, Wako, Saitama 305-0198, Japan}%
\author{H.~Watanabe}%
\affiliation{School of Physics, and International Research Center for Nuclei
  and Particles in the Cosmos, Beihang University, Beijing 100191, China}%
\affiliation{RIKEN Nishina Center for Accelerator-Based Science, Wako, Saitama 305-0198, Japan}%
\affiliation{Wako Nuclear Science Center, Institute of Particle and Nuclear Studies, High Energy Accelerator Research Organization (KEK), Wako, Saitama 351-0198, Japan}%
\author{G.J.~Lane}%
\affiliation{Department of Nuclear Physics, Research School of Physics,
  Australian National University, Canberra, ACT 2601, Australia}%
\author{M.~Ahmed}%
\altaffiliation[Present address: ]{Department of Physics, Bangladesh University of Textiles}%
\affiliation{University of Tsukuba, Tsukuba, Ibaraki 305-0006, Japan}%
\affiliation{Wako Nuclear Science Center, Institute of Particle and Nuclear Studies, High Energy Accelerator Research Organization (KEK), Wako, Saitama 351-0198, Japan}%
\author{M.~Brunet}%
\affiliation{Department of Physics, University of Surrey, Guildford, GU2 7XH, United Kingdom}%
\author{T.~Hashimoto}%
\affiliation{Rare Isotope Science Project, Institute for Basic Science (IBS), Daejeon 305-811, Republic of Korea}%
\author{S.~Ishizawa}%
\affiliation{Graduate School of Science and Engineering, Yamagata University, Yamagata 992-8510, Japan}%
\affiliation{Wako Nuclear Science Center, Institute of Particle and Nuclear Studies, High Energy Accelerator Research Organization (KEK), Wako, Saitama 351-0198, Japan}%
\affiliation{RIKEN Nishina Center for Accelerator-Based Science, Wako, Saitama 305-0198, Japan}%
\author{S.~Kimura}%
\affiliation{RIKEN Nishina Center for Accelerator-Based Science, Wako, Saitama 305-0198, Japan}%
\author{F.G.~Kondev}%
\affiliation{Physics Division, Argonne National Laboratory, Lemont, Illinois 60439, USA}%
\author{Yu.~A.~Litvinov}%
\affiliation{GSI Helmholtzzentrum f{\"{u}}r Schwerionenforschung, 64291 Darmstadt, Germany}%
\author{H.~Miyatake}%
\affiliation{Wako Nuclear Science Center, Institute of Particle and Nuclear Studies, High Energy Accelerator Research Organization (KEK), Wako, Saitama 351-0198, Japan}%
\author{J.Y.~Moon}%
\affiliation{Rare Isotope Science Project, Institute for Basic Science (IBS), Daejeon 305-811, Republic of Korea}%
\author{T.~Niwase}%
\affiliation{Department of Physics, Kyushu University, Nishi-ku, Fukuoka 819-0395, Japan}%
\affiliation{Wako Nuclear Science Center, Institute of Particle and Nuclear Studies, High Energy Accelerator Research Organization (KEK), Wako, Saitama 351-0198, Japan}%
\affiliation{RIKEN Nishina Center for Accelerator-Based Science, Wako, Saitama 305-0198, Japan}%
\author{M.~Oyaizu}%
\affiliation{Wako Nuclear Science Center, Institute of Particle and Nuclear Studies, High Energy Accelerator Research Organization (KEK), Wako, Saitama 351-0198, Japan}%
\author{J.H.~Park}%
\altaffiliation[Present address: ]{Advanced Radiation Technology Institute, Korea Atomic Energy Research Institute}%
\affiliation{Rare Isotope Science Project, Institute for Basic Science (IBS), Daejeon 305-811, Republic of Korea}%
\author{Zs.~Podoly\'{a}k}%
\affiliation{Department of Physics, University of Surrey, Guildford, GU2 7XH, United Kingdom}%
\author{M.~Rosenbusch}%
\affiliation{Wako Nuclear Science Center, Institute of Particle and Nuclear Studies, High Energy Accelerator Research Organization (KEK), Wako, Saitama 351-0198, Japan}%
\author{P.~Schury}%
\affiliation{Wako Nuclear Science Center, Institute of Particle and Nuclear Studies, High Energy Accelerator Research Organization (KEK), Wako, Saitama 351-0198, Japan}%
\author{M.~Wada}%
\affiliation{Wako Nuclear Science Center, Institute of Particle and Nuclear Studies, High Energy Accelerator Research Organization (KEK), Wako, Saitama 351-0198, Japan}%
%
%


\begin{abstract}
  De-excitation $\gamma $ rays associated with an isomeric state
  of $^{186}$Ta were investigated.
  The isomers were produced
  in multinucleon transfer reactions between a $^{136}$Xe beam and
  a natural W target, and were collected and separated by the KEK Isotope
  Separation System.
  Two $\gamma $ transitions with energies of 161.1(2) and 186.8(1)~keV
  associated with an isomeric decay
  were observed for the first time.
  The half-life of the isomeric state of the neutral atom $^{186 \rm m}$Ta
  was deduced as 17(2)~s.
  Based on the comparison with the previous measurements of
  the isomeric state using the ESR storage ring at GSI Darmstadt
  and the coupling of angular momenta of individual particle orbitals
  in odd-odd nuclei,
  a decay scheme of $^{186 \rm m}$Ta was proposed.
\end{abstract}
%
\maketitle
%
%
\section{\label{sec:lavel1}Introduction}%
Heavy neutron-rich nuclei with mass number ($A$) between 180 and 200
reveal a variety of properties
concerning their complex nuclear structures.
Shape transitions from axially symmetric prolate shapes
to oblate shapes through triaxial $\gamma $ softness
are known to take place along isotopic chains from the neutron midshell
as one approaches the closed shell at $N=126$ due to
intertwining single-particle orbitals
in a deformed nuclear potential~\cite{nom11,rob09,sar08,ste05_2}.
High-$K$ isomeric states characteristic of this nuclear region arise
from the occupation of
specific single-particle orbitals
with the large angular momentum projection
on the symmetry axis of the axial shape~\cite{dra16,wal01}.
Experimental studies of $K$ isomers and nuclear deformations
in this nuclear region are essential to enlighten
the interplay between single-particle and collective degrees of freedom
which formulate midshell nuclear structures.
In particular, odd-odd deformed nuclei, which have complex nuclear structures
related to proton-neutron interactions,
sometimes exhibit long half-lives of isomeric transitions~\cite{jai98}.
For example, the isomeric state of $^{180}_{~73}$Ta$_{107}$ at an excitation
energy of 77~keV has a half-life of $>4.5\times 10^{16}$~yr
which is much longer than the ground-state half-life, 8.15 h~\cite{leh17}.
In spite of such interesting features of odd-odd nuclei in this mass region,
experimental data of neutron-rich odd-odd nuclei
are scarce in addition to their theoretical difficulties.
Because most elements in the region are refractory,
those neutron-rich isotopes are difficult to provide  as a low-energy
beam for their nuclear spectroscopy after their production.
Recently, the KEK Isotope Separation System (KISS)~\cite{hir15} provides
opportunities to perform spectroscopic studies of those neutron-rich
refractory isotopes
using multinucleon transfer (MNT) reactions~\cite{wat15}
and in-gas laser ionization~\cite{kud96}.

The $\beta $-decay properties of the ground state
in the neutron-rich odd-odd nucleus $^{186}$Ta
were intensively investigated~\cite{mon69,pat70,guj73}
after its discovery in 1955~\cite{poe55}.
However, the information on its excited states was unknown for a long time.
Xu {\it et al.} reported the identification of
an isomeric state of $^{186}$Ta in 2004,
which disintegrates to $^{186}$W via a $\beta $ decay with
a half-life of 1.54(5)~min~\cite{xu04}.
Recently, the excitation energy of an isomeric state of $^{186}$Ta
was measured by the Experimental Storage Ring (ESR)
at GSI Darmstadt~\cite{ree12}.
They found the isomeric state at an energy of 336(20)~keV,
and observed five $\gamma $-decay and three
$\beta $-decay or internal-conversion (IC) events
with a half-life of $3.0^{+1.5}_{-0.8}$~min for the
hydrogen-like $^{186\rm m}$Ta$^{72+}$ ion.
Recent theoretical studies concerning the configuration
of isomers in $^{186}$Ta using a two-quasiparticle rotor model (TQRM)
indicates the assignment of $K^{\pi }=5^{-}\{\pi 7/2^{+}[404]\otimes \nu 3/2^{-}[512]\}$,
$2^{-}\{\pi 7/2^{+}[404]\otimes \nu 3/2^{-}[512]\}$, and
$8^{-}\{\pi 7/2^{+}[404]\otimes \nu 9/2^{-}[505]\}$
for the ground state and two isomeric states with the half-lives
of 1.54~min and 3.0~min, respectively~\cite{soo14}.
In spite of these efforts to understand the isomeric states
both in experimental and theoretical studies,
they have not been identified yet
in terms of their energies, spins and parities
by direct measurements of isomeric decays.
Presently we report a new $\gamma $-ray measurement
in coincidence with $\beta $ rays and conversion electrons
for $^{186}$Ta produced by MNT reactions, which has identified
a single isomeric state and its decay scheme.
\section{\label{sec:lavel1}Experiment}%
The experiment was
performed using the KISS at the RIKEN RIBF facility, Japan.
It focused on the isomeric decay of $^{186,187}$Ta,
and the properties of $^{187}$Ta revealed through its isomeric decay
were reported in Ref.~\cite{wal20_2}.
$^{136}$Xe beams accelerated up to 7.2~MeV/nucleon by the
RIKEN Ring Cyclotron (RRC) accelerator
were incident on a natural tungsten target of thickness
5~$\mu $m attached to a rotating wheel.
The typical beam intensity was 50~pnA on the target.
Various nuclei produced by the MNT reactions
between $^{136}$Xe and $^{\rm nat}$W were ejected from the target
and passed through
a polyimide film of thickness 5~$\mu $m
into a doughnut-shaped gas cell~\cite{hir17},
which was filled with high-purity argon gas of pressure 80~kPa.
After thermalization and neutralization in the argon gas,
the now neutral atoms were transported to the exit of the gas cell.
They were irradiated with two-color lasers for element-selective
ionization with the laser resonance ionization technique
just before the exit of the gas cell.
The ionization scheme of tantalum was investigated before the
experiment~\cite{hir19}.
The ions ejected from the gas cell
were transported by a stack of three multi-pole RF ion guides,
and were accelerated through a voltage of 20~kV.
The ions with $A=186$ were mass-separated
by a dipole electromagnet with a resolving power of $A/\Delta A=900$,
and were implanted into an aluminized Mylar tape of thickness
12.5~$\mu $m.

The Multi-Segmented Proportional Gas Counter (MSPGC)~\cite{muk18,hir21}
was placed surrounding the tape to detect $\beta $ rays,
X rays and conversion electrons produced in the decay of
the implanted radioactive nuclei.
The MSPGC consists of two concentric layers
with 16 proportional gas counter tubes in each layer.
The geometry is 200~mm in height and 90~mm in outer diameter.
Four High-Purity Germanium (HPGe) clover detectors were placed
surrounding the MSPGC to detect $\gamma $ rays.
With the compact configuration of the experimental setup,
where the distances between the surfaces of
the HPGe clover detectors and the tape were around 5~cm,
the total absolute detection efficiency for full-energy peaks was
15\% for 150-keV $\gamma $ rays.
%
The beams from KISS were pulsed by an electrostatic deflector after
the dipole electromagnet.
Two kinds of time cycles were used; one was 1800-s beam-on and 1800-s beam-off
periods (long cycle),
and the other was 300-s beam-on and 300-s beam-off periods (short cycle).
The tape was moved vertically by about 30~cm
after each cycle to eliminate radioactivities from
both the preceding implantation and the accumulated daughter nuclei.
The data were accumulated in 3 long cycles and 71 short cycles
with $^{186}$Ta beams of around 5 pps from the KISS.
\section{\label{sec:lavel1}Analysis and results}%
A hit pattern analysis of the 32 gas counter tubes
in the MSPGC makes it possible
to separate different kinds of events.
The hit pattern ``$M=2$'', where one telescope (a pair of both inner
and outer counter tubes on the same radius vector)
fires, is sensitive to energetic $\beta $ rays.
\begin{figure}%
  \includegraphics[width=86mm]{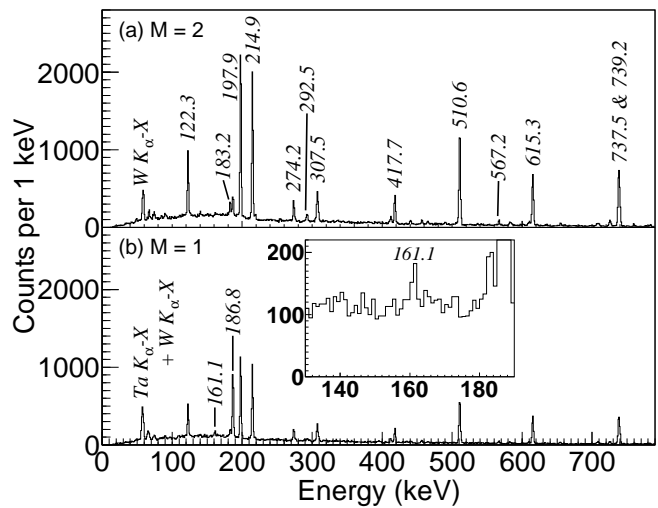}%
  \caption{%
    $\gamma $-ray energy spectra in coincidence with
    the MSPGC hit patterns (a) ``$M=2$'' and (b) ``$M=1$''.
    \label{fig:spectra}%
  }%
\end{figure}%
Figure~\ref{fig:spectra} (a) indicates the $\gamma $-ray energy spectrum
in coincidence with the MSPGC hit pattern ``$M=2$''
summed for long- and short-cycle runs.
A peak corresponding to the K$_{\alpha }$ X rays of
W indicates that the transitions in the daughter nuclei
of the $^{186}$Ta $\beta $ decay contribute
when this hit pattern condition is applied.
Twelve peaks labeled by energy values
correspond to the known $\beta $-delayed transitions
of $^{186}$Ta~\cite{A186_2003}.
\begin{figure}%
  \includegraphics[width=86mm]{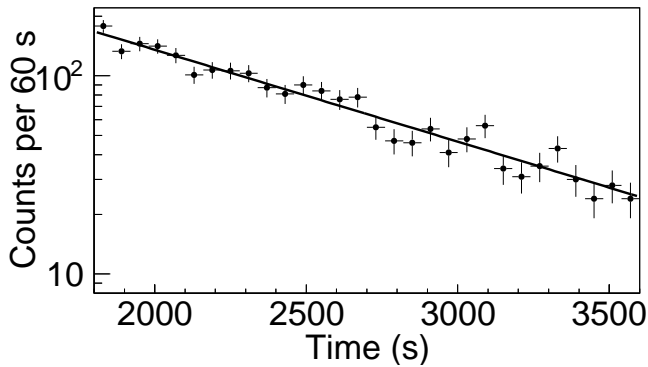}%
  \caption{%
    $\gamma $-ray time spectrum summed for twelve transitions
    labeled in Fig~\ref{fig:spectra} (a)
    in coincidence with the MSPGC hit pattern ``$M=2$''.
    The line through the data is a log-likelihood fit to the decay (beam off)
    period from 1800 to 3600~s, yielding
    a half-life of 10.8(5)~min.
    \label{fig:decay_ground}%
  }%
\end{figure}%
Figure~\ref{fig:decay_ground} shows a $\gamma $-ray time spectrum summed
for those twelve transitions
in coincidence with the MSPGC hit pattern ``$M=2$''
for the long-cycle run.
The half-life was deduced by fitting the time spectrum
to a combined function of a
decay curve and a constant background for the decay period from 1800 to 3600~s;
the fit is shown as a solid line.
The half-life obtained in the fitting, with a reduced $\chi ^{2}=1.09$,
is 10.8(5)~min, which agrees with the literature value
of the $^{186}$Ta ground-state half-life, 10.5(3)~min~\cite{A186_2003}.
We could find no evidence for a decay component with a half-life
around 1.5~min reported in Ref.~\cite{xu04}.

Figure~\ref{fig:spectra} (b) exhibits the $\gamma $-ray energy spectrum
in coincidence with the MSPGC hit pattern ``$M=1$'',
where only one counter tube in the inner layer
fires. This detection mode
is sensitive to X rays and low-energy conversion electrons
by suppressing the events owing to the energetic $\beta $ rays.
It is found that the peak heights corresponding to the $\beta $-delayed
$\gamma $ rays decrease.
Furthermore, the emergence of a peak corresponding to the K$_{\alpha }$ X rays of
Ta indicates that the conversion electrons fire the MSPGC
with this hit pattern condition.
Two previously unreported peaks at energies of 161.1(2)~keV and 186.8(1)~keV
were found.
The inset enlarges the spectrum at the energies around 160~keV.

\begin{figure}%
  \includegraphics[width=86mm]{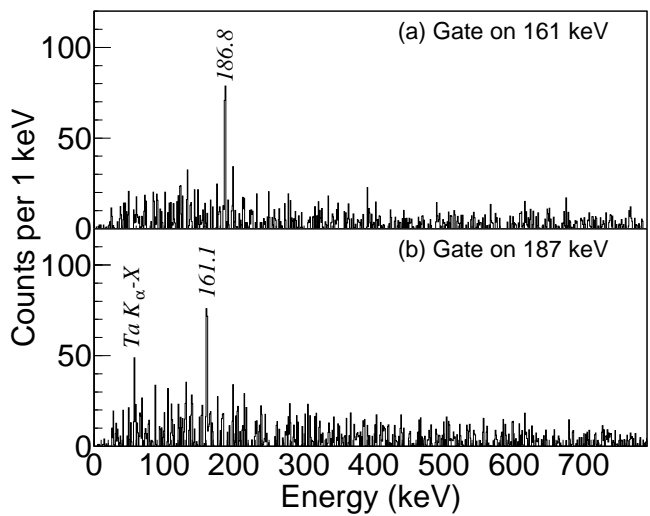}%
  \caption{%
    Backgound-subtracted $\gamma $-ray energy projections
    measured with gates on
    (a) 161- and (b) 187-keV $\gamma $ rays, respectively,
    and without MSPGC coincidence.
  }%
  \label{fig:gg-coin}%
\end{figure}%
Figure~\ref{fig:gg-coin} (a) and (b) show background-subtracted
$\gamma $-ray energy projections measured
with gates on the 161- and 187-keV $\gamma $ rays,
respectively,
and without MSPGC coincidence.
They clearly prove the coincidence between the 187- and 161-keV
$\gamma $ rays, indicating that they are cascading transitions.
The counts of the 161- and 187-keV $\gamma $ peaks
in Fig.~\ref{fig:spectra} (b)
are 147(32) and 1845(50), respectively.
Because the 161-keV peak count is more than ten times smaller
than the 187-keV peak count, the 161-keV transition is considered
to be highly converted implying a large multipolarity,
and further that the 161-keV transition is the isomeric transition.
It is supported by the K$_{\alpha }$-X-ray peak of Ta
shown in the $\gamma $-ray energy spectrum
in coincidence with the 187-keV $\gamma $ rays
on Fig.~\ref{fig:gg-coin} (b).

\begin{figure}%
  \includegraphics[width=86mm]{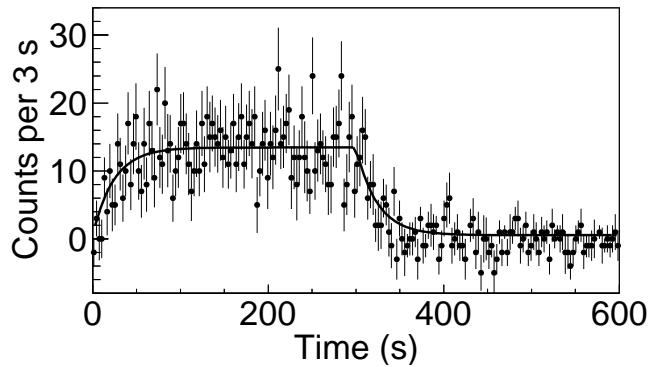}%
  \caption{%
    Background-subtracted
    $\gamma $-ray time spectrum summed for 161- and 187-keV transitions
    in coincidence with the MSPGC hit pattern ``$M=1$''.
    The line through the data is a log-likelihood fit to the growth (beam on)
    and decay (beam off) periods, each of 300~s duration, yielding
    a half-life of 17(2)~s.
    \label{fig:decay}%
  }%
\end{figure}%
Figure~\ref{fig:decay} shows a background-subtracted
$\gamma $-ray time spectrum summed
for 161- and 187-keV transitions in coincidence
with the MSPGC hit pattern ``$M=1$'' for the short-cycle runs.
The half-life was deduced by fitting the time spectrum to
a combined function of a
growth curve, a decay curve and a constant background;
the fit is shown as a solid line.
The half-life obtained in the fitting, with a reduced $\chi ^{2}=0.94$,
is 17(2)~s, which is shorter than the half-life
of the ground state of $^{186}$Ta, 10.5(3)~min~\cite{A186_2003}.
Therefore the newly found 161- and 187-keV
$\gamma $ transitions are considered to be cascading trasitions
from an isomeric state in $^{186}$Ta,
with a half-life of 17(2)~s.

\begin{figure}%
  \includegraphics[width=86mm]{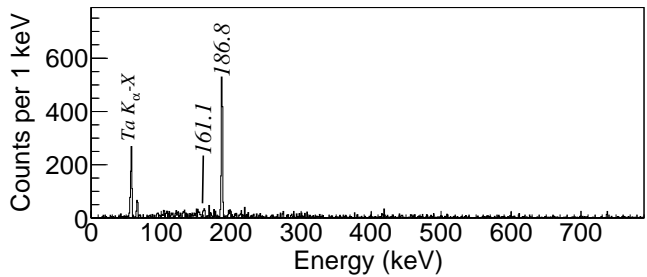}%
  \caption{%
    $\gamma $-ray energy spectrum with a time gate from 100~s to 300~s
    subtracting a spectrum with a time gate from 400~s to 600~s
    in the short cycle runs with the coincidence of the MSPG hit pattern
    ``$M=1$'' and ``$M=2$''.
    \label{fig:time_gate_spectrum}%
  }%
\end{figure}%
To investigate a possible $\beta $-decay mode from the isomeric state,
a $\gamma $-ray energy spectrum with a time gate from 400~s to 600~s
was subtracted from a spectrum with a time gate from 100~s to 300~s
in the short cycle runs with the coincidence of the MSPG hit pattern
``$M=1$'' and ``$M=2$'' as shown in Fig.~\ref{fig:time_gate_spectrum}.
Because there are no $\gamma $-ray peaks that follow
a decay curve with the half-life of 17~s,
except for the 161- and 187-keV lines,
the isomeric decay predominantly proceeds
either with $\gamma $ transition or with IC.

The energy of the isomeric state, 347.9(2)~keV,
was obtained from the sum of the two transition energies.
It agrees with the excitation energy
measured by the ESR at GSI Darmstadt, 336(20)~keV~\cite{ree12}.
\section{\label{sec:lavel1}Discussion}%
The obtained half-life of the neutral $^{186\rm m}$Ta,
17(2)~s, is shorter than
the half-life measured in the ESR,
$3.0^{+1.5}_{-0.8}$~min, for a hydrogen-like
$^{186\rm m}$Ta$^{72+}$~\cite{ree12}.
The difference is considered to come from
the components of IC, which are suppressed
in the hydrogen-like ion.
The ratio of the half-lives, $0.09^{+0.03}_{-0.05}$,
should be compared to $(1+\alpha _{{\rm K}(1e)})/(1+\alpha_{\rm tot})$,
where $\alpha _{\rm tot}$ and $\alpha _{{\rm K}(1e)}$ are
the total conversion coefficient for the neutral atom
and the K-conversion coefficient for the hydrogen-like ion, respectively.
\begin{table}
  \caption{\label{tab:conversion_161kev}%
    Theoretical values of total and K conversion coefficients
    in the neutral atom,
    $\alpha _{\rm tot}$ and $\alpha _{\rm K}$~\cite{kib08}, respectively,
    and in the hydrogen-like ion, $\alpha _{{\rm K}(1e)}$~\cite{kib_pc},
    for the 161-keV transition in $^{186}$Ta.
    The last column indicates the ratio of $1+\alpha _{{\rm K}(1e)}$
    to $1+\alpha _{\rm tot}$.}%
\begin{ruledtabular}
\begin{tabular}{crrrr}%
  Multipolarity&\multicolumn{1}{c}{$\alpha _{\rm tot}$}&\multicolumn{1}{c}{$\alpha _{{\rm K}}$}&\multicolumn{1}{c}{$\alpha _{{\rm K}(1e)}$}&\multicolumn{1}{c}{$\frac{1+\alpha _{{\rm K}(1e)}}{1+\alpha _{\rm tot}}$}\\
  \hline
  E1&0.11 &0.09 &0.044 &0.94\\
  M1&1.1  &0.9  &0.47  &0.70\\
  E2&0.6  &0.3  &0.15  &0.72\\
  M2&6.9  &5.2  &2.6   &0.46\\
  E3&6.3  &0.9  &0.44  &0.20\\
  M3&36   &20   &10    &0.30\\
\end{tabular}
\end{ruledtabular}
\end{table}
Table~\ref{tab:conversion_161kev} summarizes those calculated
conversion coefficients of the 161-keV transition of $^{186}$Ta
for various multipolarities.
The value of $(1+\alpha _{{\rm K}(1e)})/(1+\alpha _{\rm tot})=0.20$
for the multipolarity E3 is the closest
to the ratio of the measured half-lives.
Furthermore, the ratio of the efficiency-corrected peak count
of Ta ${\rm K}_{\alpha }$ X rays to the 161-keV $\gamma $ rays
in Fig.~\ref{fig:gg-coin} (b) is 0.70(18).
It should be compared to $\alpha _{\rm K}$.
The value of $\alpha _{\rm K}\times \omega_{\rm K}/\left\{1+p({\rm K}_{\beta })/p({\rm K}_{\alpha })\right\}=0.68$ for the multipolarity E3, where
$\omega _{\rm K}=0.952(4)$ and $p({\rm K}_{\beta })/p({\rm K}_{\alpha })=0.267(4)$~\cite{sch96}
are the K-shell fluorescence yield and the emission probability ratio
of Ta, respectively, 
also agrees with the measured value.
Therefore, the transition de-exciting the 17-s isomer is interpreted as having
a multipolarity of E3.

The measurements in the ESR reported five $\gamma $-decay and three
$\beta $-decay or IC events
from the isomeric state~\cite{ree12}.
It should be noted that the $\beta $-decay and IC events
cannot be distinguished in the ESR data.
When the five $\gamma $-decay events are by the E3 transition
with $\alpha _{{\rm K}(1e)}=0.44$,
2.2(10) events are expected by IC in average.
Observation of three $\beta $-decay or IC events in the ESR
support the dominance of $\gamma $-decay and IC modes
from the isomeric state.

In Fig.~\ref{fig:gg-coin}, the 187-keV (161-keV) $\gamma $ ray is observed
when it is detected by
the HPGe detectors
in coincidence with
a conversion electron from the 161-keV (187-keV) transition
detected by the MSPGC.
The detected counts of the 161- and 187-keV $\gamma $ rays,
$N_{161}$ and $N_{187}$,
are written as,
\begin{eqnarray}%
  N_{161}=&
  N_{\rm iso}\times \frac{1}{1+\alpha _{161}}\times \varepsilon _{\gamma 161}\times \frac{\alpha _{187}}{1+\alpha _{187}}\times \varepsilon _{{\rm CE} 187},\\
  N_{187}=
  &N_{\rm iso}\times \frac{1}{1+\alpha _{187}}\times \varepsilon _{\gamma 187}\times \frac{\alpha _{161}}{1+\alpha _{161}}\times \varepsilon _{{\rm CE} 161},
\end{eqnarray}%
where $N_{\rm iso}$ is the number of isomeric decays,
and $\alpha _{161(187)}$, $\varepsilon _{\gamma 161(187)}$ and
$\varepsilon _{{\rm CE}161(187)}$ are the conversion coefficient,
the $\gamma $-ray full-energy peak efficiency
and the conversion-electron detection
efficiency for the 161(187)-keV transition.
Therefore the ratio of conversion coefficients for the 161- and
187-keV transitions is related to the ratio of
their detected $\gamma $-ray counts through the equation,
\begin{equation}%
\frac{\alpha _{161}}{\alpha _{187}}
=\frac{N_{187}}{N_{161}}\times \frac{\varepsilon _{\gamma 161}}{\varepsilon _{\gamma 187}},
\end{equation}%
where ${\varepsilon _{{\rm CE}187}}={\varepsilon _{{\rm CE}161}}$
is assumed.
\begin{table}
  \caption{\label{tab:conversion_187kev}%
    Calculated conversion coefficients
    for the 187-keV transition of $^{186}$Ta,
    $\alpha (187~{\rm keV})$~\cite{kib08},
     and the ratio compared to the
  conversion coefficient of the E3 161-keV transition.}
\begin{ruledtabular}
\begin{tabular}{crr}%
  Multipolarity&\multicolumn{1}{c}{$\alpha (187~{\rm keV})$}&\multicolumn{1}{c}{$\frac{\alpha (161~{\rm keV}, {\rm E3})}{\alpha (187~{\rm keV})}$}\\
  \hline
  E1&0.072&88\\
  M1&0.74&8.5\\
  E2&0.37&17\\
  M2&4.0 &1.6\\
\end{tabular}
\end{ruledtabular}
\end{table}
Table~\ref{tab:conversion_187kev} summarizes the calculated conversion
coefficients of the 187-keV transition for various electromagnetic
multipolarities~\cite{kib08}.
The last column indicates the ratio of the conversion coefficients
for an E3 161-keV transition compared to a 187-keV transition.
The ratio of the efficiency-corrected $\gamma $-ray count for the 187-keV
transition to the 161-keV transition is 12.8(28),
which agrees with ratios for the pure M1 and E2 transitions
in Table~\ref{tab:conversion_187kev}
within two standard deviations.
The 187-keV transition can most probably be considered as
a mixed M1/E2 transition.
Assuming the theoretical conversion coefficient of the E3 161-keV transition
in Table~\ref{tab:conversion_161kev}, $\alpha_{161}=6.3$,
the conversion coefficient of
the M1/E2-mixed 187-keV transition becomes
$\alpha({\rm M}1/{\rm E}2) =0.49(11)$
from the measured $\gamma $-ray intensity ratio, 12.8(28).
%

%
\begin{figure}%
  \includegraphics[width=86mm]{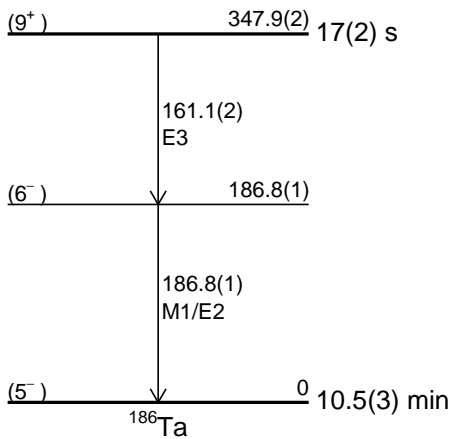}%
  \caption{%
    Proposed decay scheme for the isomeric state in $^{186}$Ta.
    Labels with arrows indicate the $\gamma $-ray energies in keV
    and the electromagnetic multipolarities are also shown.
    The ground-state half-life is from Ref.~\cite{A186_2003}.
    \label{fig:scheme}%
  }%
\end{figure}%
Based on the above considerations,
we propose a decay scheme of the isomeric state
as shown in Fig.~\ref{fig:scheme}.
The spin and parity of the ground state
for the nearest neighbor odd-$A$ isotope and isotone of $^{186}$Ta,
$^{185}$Ta and $^{187}$W,
is $(7/2^{+})$ and $3/2^{-}$, respectively~\cite{A185_2005,A187_2009}.
They are considered to be due to the single-particle orbitals
of the 73rd proton, $\pi 7/2^{+}[404]$,
and the 113th neutron, $\nu 3/2^{-}[512]$, respectively.
The coupling of the angular momenta of those individual particle states
in $^{186}$Ta gives the lower energy state with
the parallel-spin configuration
$K^{\pi }=5^{-}\{\pi 7/2^{+}[404]\otimes \nu 3/2^{-}[512]\}$~\cite{gal58,soo14},
which is considered as the ground state of $^{186}$Ta.
We assigned $6^{-}$ and $9^{+}$ for the 187-keV excited state
and the isomeric state, respectively,
based on $5^{-}$ of the ground state
and the multipolarities discussed above.
The proposed decay scheme indicates that the $K^{\pi }=9^{+}$
isomeric state decays to the first member of the
rotational band of the $K^{\pi }=5^{-}$
ground state.
Sood and Gowrishankar suggested $K^{\pi }=8^{-}$
for the isomeric state~\cite{soo14},
however our measurements indicate $K^{\pi }=9^{+}$.
They also suggest that a $K^{\pi }=9^{+}$ state is possible
with a antiparallel-spin
two-quasiparticle configuration $\pi 7/2[404]\otimes \nu 11/2[615]$.
The parallel-spin partner of the isomeric state,
$K^{\pi }=2^{+}\{\pi 7/2^{+}[404]\otimes \nu 11/2^{+}[615]\}$,
and the antiparallel-spin partner of the ground state,
$K^{\pi }=2^{-}\{\pi 7/2^{+}[404]\otimes \nu 3/2^{-}[512]\}$,
are located at excited energies
between the isomeric state and the ground state~\cite{gal58,soo14}.
However the transition from the isomeric state to those states is
strongly suppressed by the large $K$ change $\Delta K=7$.

Adopting the theoretical conversion coefficient 6.3
for the E3 161-keV transition in Table~\ref{tab:conversion_161kev},
its partial $\gamma $-decay half-life,
$T_{1/2}^{\gamma }$, becomes
2.1(2)~min, which gives a reduced transition strength
of $B({\rm E}3)=1.7(2)\times 10^{-3}~{\rm W.u.}$
The hindrance factor $F_{\rm W}=T_{1/2}^{\gamma }/T_{1/2}^{\rm W}$,
where $T_{1/2}^{\rm W}$ is the Weisskopf estimate of the half-life,
is obtained as $5.9(7)\times 10^{2}$.
This value is consistent with the experimental range for
E3 transitions in Ref.~\cite{lob68,kon15}
with $K$ forbiddeness $\nu =\Delta K-\lambda =1$,
where $\Delta K=4$ and
the transition multipolarity is $\lambda =3$.
The direct M4 348-keV transition from the isomeric state
to the ground state with $\Delta K=4$ and $\lambda =4$
could have a lower hindrance factor
$F_{\rm W}\sim 10^{-1}$~\cite{lob68}.
However, the corresponding partial half-life, 20~min, obtained by considering
the theoretical conversion coefficient, 5.3~\cite{kib08},
is much longer than the measured half-life, 17(2)~s,
indicating that such a direct M4 transition is significantly
suppressed.
\section{\label{sec:lavel1}Summary}%
We have measured $\gamma $ rays associated
with the isomeric decay of a 348-keV state in  $^{186}$Ta,
and found two transitions with energies of 161.1(2)
and 186.8(1)~keV for the first time.
The fit to the time spectrum for those two transitions
indicates an isomeric state with a half-life of 17(2)~s.
The comparison with the previous measurement obtained with the ESR at GSI
Darmstadt, suggests that the 161-keV isomeric transition
has an electromagnetic multipolarity of E3,
which is followed by the 187-keV transition (M1/E2).
A decay scheme of $^{186 \rm m}$Ta was proposed
by considering
the coupling of the angular momenta of individual
particle orbitals in odd-odd nuclei,
assigning spin-parities of $9^{+}$, $6^{-}$ and $5^{-}$
to the isomeric, excited and ground states, respectively.
It indicates that the $K^{\pi }=9^{+}$ isomer decays to the first member
in the rotational band of the $K^{\pi }=5^{-}$ ground state.
The hindrance factor for the isomeric transition
is consistent with systematics~\cite{lob68,kon15}
supporting the proposed
decay scheme.
We could not find any evidence for $\beta $ decay
from the observed isomer,
or from any other state with a half-life around 1.5~min~\cite{xu04},
in our measurements.
\begin{acknowledgments}%
This experiment was performed at RI Beam Factory operated
by RIKEN Nishina Center and CNS, University of Tokyo.
The authors gratefully acknowledge the accelerator staff
for their support. This work was funded in part by Grants
No. JP23244060, No. JP24740180,
No. JP26247044, No. JP15H02096, No. JP17H01132, No. JP17H06090,
and No. JP18H03711 from JSPS KAKENHI;
No. ST/P005314/1 from United Kingdom STFC;
No. 11921006 and No. 11835001 from NSFC;
No. 682841 ``ASTRUm'' from ERC (Horizon 2020);
and No. DE-AC02-06CH11357 from U.S. Department of Energy (Office of Nuclear
Physics).
\end{acknowledgments}%

\bibliography{paper}

\end{document}